\documentclass[sn-nature]{sn-jnl}

\usepackage{graphicx}%
\usepackage{multirow}%
\usepackage{amsmath,amssymb,amsfonts}%
\usepackage{amsthm}%
\usepackage{mathrsfs}%
\usepackage[title]{appendix}%
\usepackage{xcolor}%
\usepackage{textcomp}%
\usepackage{manyfoot}%
\usepackage{booktabs}%
\usepackage{algorithm}%
\usepackage{algorithmicx}%
\usepackage{algpseudocode}%
\usepackage{listings}%
\usepackage{wrapfig}
\usepackage{sidecap}
\theoremstyle{thmstyleone}%
\theoremstyle{thmstyletwo}%

\theoremstyle{thmstylethree}%

\newcommand{\parspace}{\vspace{0.15cm}}

\begin{document}

\vspace{-2cm}
\title[Article Title]{Revisiting Lipid Nanoparticle Composition and Structure: A Critical Take on Simulation Approaches}

\author[1,2]{\fnm{Marius F.W.} \sur{Trollmann}}\email{marius.trollmann@fau.de}
\author[1]{\fnm{Paolo} \sur{Rossetti}}\email{paolo.rossetti@fau.de}
\author*[1,2,3]{\fnm{Rainer A.} \sur{B\"ockmann}}\email{rainer.boeckmann@fau.de}
\equalcont{These authors contributed equally to this work.}

\affil[1]{Computational Biology, Department of Biology, Friedrich-Alexander-Universit\"at Erlangen-N\"urnberg, Erlangen, Germany}
\affil[2]{Erlangen National High-Perfomance Computing Center (NHR@FAU), Erlangen, Germany}
\affil[3]{FAU Profile Center Immunomedicine (FAU I-MED), Erlangen, Germany}
\affil[*]{rainer.boeckmann@fau.de}

\vspace{-2cm}
\maketitle

\begin{figure}
\centering
\includegraphics[width=12cm]{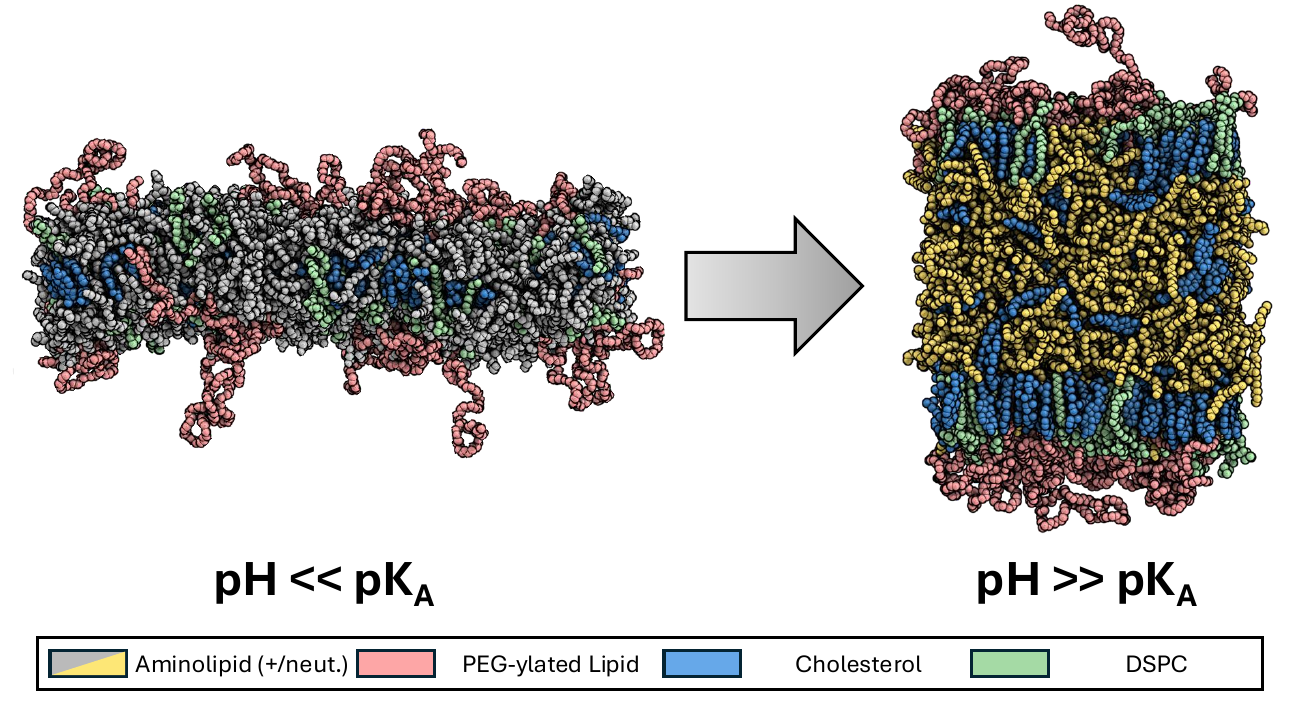}
\caption{Equilibrated LNP formulations at low and high pH~\cite{Trollmann2022-aw}. At low pH, the protonated aminolipids from, together with the helper lipids, a lipid bilayer. Upon increase in pH, here a deprotonation of the aminolipids, a phase transition occurs that is characterized by a drastic decrease of the lateral membrane area and the formation of a hydrophobic core. This transition is artificially blocked for simulations within the $NP_zT$ ensemble~\cite{Garaizar2024-ts}.}
\label{fig:structure}
\end{figure}

The impact of pH on the structural organization of lipid nanoparticles (LNPs) has been well-documented, with both all-atom and coarse-grained simulations revealing a pH-dependent phase transition~\cite{Trollmann2022-aw, Philipp2023-hq, Paloncyova2023-kv, Ramezanpour2019-yx}. Specifically, at low pH (where aminolipids are fully protonated), LNP-mimetic systems exhibit a lipid bilayer phase, while at high pH, they transition to an LNP-like structure (Fig.~\ref{fig:structure}). This structure comprises a hydrophobic core primarily made up of aminolipids and cholesterol, surrounded by a monolayer of helper lipids (e.g., DSPC and PEGylated lipids), cholesterol, and a few aminolipids~\cite{Trollmann2022-aw, Paloncyova2023-kv, Yanez-Arteta2018-am, Kjolbye2024-cq}. 
\parspace

A recent study published in PNAS~\cite{Garaizar2024-ts} challenges this understanding by employing \textit{direct coexistence} (DC) simulations. Rather than simulating complete LNPs, the authors modeled smaller cross-sectional systems with a hydrophobic core sandwiched between lipid monolayers in an aqueous environment --- an approach similar to previous studies~\cite{Ramezanpour2019-yx, Trollmann2022-aw, Kjolbye2024-cq}. However, unlike prior results, these DC simulations indicated minimal structural changes between neutral and acidic pH (see Fig.~5 of~\cite{Garaizar2024-ts}), with protonated lipids remaining uniformly distributed within the core (as shown in Fig.~2B of~\cite{Garaizar2024-ts}), leading to charge densities of approximately 0.4-0.5\,e$_\text{o}$/nm$^{3}$. This charge distribution implies a significant electrostatic energy of approximately  $10^{11}\dots10^{13}$\,kJ/mol for LNPs of 100-200\,nm in size (as in the LNP1 and LNP2 systems described in~\cite{Garaizar2024-ts}, dielectric constant $\epsilon_r \approx 10$ assumed). This level of energy is roughly 20–140 times the equivalent of TNT. 
\parspace

Such a charge distribution seems improbable and may stem from the chosen simulation ensemble. By fixing the monolayer area while coupling only the $z$-direction (normal to the surface) to environmental pressure ($NP_zT$ ensemble), the study’s setup restricts the system’s ability to undergo structural modifications that would naturally arise from the protonation of all aminolipids (from a core-monolayer phase to a lipid bilayer) or from changes in LNP formulation. This constraint likely hinders an unbiased exploration of the LNP core-shell architecture and establishes an energetic barrier that inhibits RNA escape under neutral pH conditions, a phenomenon suggested from previous unbiased simulations~\cite{Trollmann2022-aw, Paloncyova2023-kv}. Moreover, differences observed between the two LNP formulations, such as DSPC-water micellar structures below the LNP surface, seem predetermined by the systemic bias introduced by this setup and highly sensitive to system configuration. Therefore, the simulations by Garaizar \textit{et al.}~\cite{Garaizar2024-ts} may not provide conclusive insights into the role of varying LNP compositions on LNP activity.
\parspace

\bibliography{sn-bibliography}

\end{document}